\documentstyle[aps,12pt,psfig]{revtex}

\newcommand{ \be }{\begin{equation}}
\newcommand{ \ee }{\end{equation}}
\newcommand{ \bea }{\begin{eqnarray}}
\newcommand{ \eea }{\end{eqnarray}}
\newcommand{ \la }{\langle}
\newcommand{ \ra }{\rangle}
\newcommand{ \mpx }{\langle p_{x} \rangle}
\newcommand{ \mpt }{\langle p_{t} \rangle}

\begin{document}
\normalsize

\title{ Directed Flow of Light Nuclei
in Au+Au Collisions at AGS Energies}

\author{
  J.~Barrette$^5$, R.~Bellwied$^{9}$, 
  S.~Bennett$^{9}$, R.~Bersch$^7$, P.~Braun-Munzinger$^2$, 
  W.~C.~Chang$^7$, W.~E.~Cleland$^6$, M.~Clemen$^6$, 
  J.~Cole$^4$, T.~M.~Cormier$^{9}$, 
  Y.~Dai$^5$, G.~David$^1$, J.~Dee$^7$, O.~Dietzsch$^8$, M.~Drigert$^4$,
  K.~Filimonov$^5$, S.~C.~Johnson$^7$, 
  J.~R.~Hall$^9$, T.~K.~Hemmick$^7$, N.~Herrmann$^2$, B.~Hong$^2$, 
  Y.~Kwon$^7$,
  R.~Lacasse$^5$, Q.~Li$^{9}$, T.~W.~Ludlam$^1$,
  S.~K.~Mark$^5$, R.~Matheus$^{9}$, S.~McCorkle$^1$, J.~T.~Murgatroyd$^9$,
  D.~Mi\'{s}kowiec$^2$,
  E.~O'Brien$^1$,  
  S.~Panitkin$^7$, P.~Paul$^7$, T.~Piazza$^7$, M.~Pollack$^7$, 
  C.~Pruneau$^9$, 
  M.~N.~Rao$^7$, E.~Reber$^4$, M.~Rosati$^5$, 
  N.~C.~daSilva$^8$, S.~Sedykh$^7$, U.~Sonnadara$^6$, J.~Stachel$^3$, 
  E.~M.~Takagui$^8$, 
  S.~Voloshin$^3$\footnote{On leave from Moscow Engineering Physics Institute,
     Moscow, 115409,  Russia},
  T.~B.~Vongpaseuth$^7$,
  G.~Wang$^5$, J.~P.~Wessels$^3$, C.~L.~Woody$^1$, 
  N.~Xu$^7$,
  Y.~Zhang$^7$, C.~Zou$^7$\\ 
(E877 Collaboration)
}
\address{
 $^1$ Brookhaven National Laboratory, Upton, NY 11973\\
 $^2$ Gesellschaft f\"ur Schwerionenforschung, 64291 Darmstadt, Germany\\
 $^3$ Universit\"at Heidelberg, 69120 Heidelberg, Germany\\
 $^4$ Idaho National Engineering Laboratory, Idaho Falls, ID 83402\\
 $^5$ McGill University, Montreal, Canada\\
 $^6$ University of Pittsburgh, Pittsburgh, PA 15260\\
 $^7$ SUNY, Stony Brook, NY 11794\\
 $^8$ University of S\~ao Paulo, Brazil\\
 $^9$ Wayne State University, Detroit, MI 48202\\
}

\date{July 31, 1998}

\maketitle

\begin{abstract}
Directed flow of deuterons, tritons, $^3$He, and $^4$He is studied in
Au+Au collisions at a beam momentum of 10.8 $A$ GeV/c.  Flow of all
particles is analyzed as a function of transverse momentum
for different centralities of the collision.  The directed flow signal,
$v_1(p_t)$, is found to increase with particle mass.  This mass
dependence is strongest in the projectile rapidity region.

\end{abstract}
\narrowtext


\section{Introduction}

Anisotropies in the azimuthal distribution of particles, also called
anisotropic (directed, elliptic, etc.) transverse flow play an important
role in high energy nuclear collisions~\cite{lqm97,ritt98}.  In
particular, flow of composite particles such as light nuclei is of a
large interest for the understanding of the nuclear collision
dynamics~\cite{gutb89,koch90,stoc94,mati95}.  At lower beam energies
(kinetic energies from about 0.2 to 1.15~GeV per nucleon) flow of light
fragments has been studied intensively both experimentally
\cite{doss87,gust88,ogil89,sull90,wang95,part95,huan96,cro97} and
theoretically~\cite{gutb89,koch90,stoc94,mati95,dani95} (and references
therein).  A theoretically predicted increase of $\mpx /A$, the mean
transverse momentum per fragment nucleon projected onto the reaction
plane, with particle mass was later confirmed
experimentally~\cite{doss87,gust88,ogil89,sull90,wang95,part95}.
Initially, the theoretical prediction of the effect was based on the
observation that the collective motion of the higher mass fragments
should be less sensitive to thermal distortions.  Later,
calculations~\cite{koch90,stoc94,mati95} showed the same effect in
models based on a picture where the production of light nuclei takes
place by the coalescence of nucleons close to each other in momentum and
configuration space.  Experimental data~\cite{wang95} for particles with
transverse momentum greater than 0.2~GeV/c per fragment nucleon were
found to be consistent with momentum space coalescence, but a complete
understanding of the effect is still lacking.  Recently, a measurement
of light fragment directed flow in Au+Au collisions at AGS energies has
been reported by the E802 Collaboration~\cite{ahle98}.  Their conclusion
was that, in the target (pseudo)rapidity region, the maximal (as a
function of centrality) values of $\mpx/A$ do not depend on the
particle species implying a common directed flow velocity.

In the current paper, we go beyond our previous measurements of
anisotropic flow at AGS energies~\cite{l877flow1,l877flow2,l877flow3}
and present results of the analysis of directed flow of deuterons,
tritons, $^3$He, and $^4$He, detected in the E877 spectrometer in Au +
Au collisions at a beam momentum of 10.8$A$ GeV/c.  Directed flow of all
particles is analyzed as a function of the transverse momentum for
different centralities of the collision.  Triton, $^3$He, and $^4$He
flow patterns are analyzed in the beam rapidity region, where these
particles can be easily identified.  Deuteron flow
is analyzed also as a function of rapidity, in the rapidity region of
2.2$<y<$3.4.  The results of a very similar analysis of proton and
charged pion flow were reported in~\cite{l877flow3} and many technical
details relevant for the current analysis can be found there.

\section{E877 apparatus and flow analysis}

The E877 apparatus is shown in Fig.~1.  In the E877 setup, charged
particles, emitted in the forward direction and passing though a
collimator ($ -134 < \theta_{horizontal} < 16 $ mrad, $ -11 <
\theta_{vertical} < 11 $ mrad), are analyzed by a high resolution
magnetic spectrometer.  The spectrometer acceptance covers mostly the
forward rapidity region.  The momentum of each particle is measured
using two drift chambers (DC2 and DC3, position resolution about
300~$\mu$m) whose pattern recognition capability is aided by four
multi-wire proportional chambers (MWPC).  The average momentum
resolution is $\Delta p/p \approx$3\% limited by multiple scattering.  A
time-of-flight hodoscope (TOFU) located directly behind the tracking
chambers provides the time-of-flight with an average resolution of
85~ps.  Energy loss in TOFU and in a Forward Scintillator array located
approximately 30~m downstream of the target is used to determine the
particle charge.

The particle identification is performed by combining measurements of
momentum, velocity, and charge of the particle.  In the present sample
of light nuclei we estimate an admixture of other particle species to be
no more than 15--20\%.  Given that particles of the admixture (mostly
other light nuclei, and some protons in the deuteron sample) exhibit a
similar flow signal (see below), we arrive at relative errors in the
final results due to particle misidentification of less than or about
5\%.

The determination of the centrality of the collision and of the reaction
plane orientation are made using the transverse energy flow measured in
the target calorimeter (TCal), and participant calorimeter (PCal).  Both
calorimeters have $2 \pi$ azimuthal coverage and, combined together,
provide nearly complete polar angle coverage: TCal and PCal cover the
pseudorapidity regions $-0.5 <\eta <0.8$ and $0.8 <\eta < 4.2$,
respectively~\cite{l877flow2}. 

The azimuthal anisotropy of particle production is studied by means of
Fourier analysis of azimuthal
distributions~\cite{lvzh,lolli,l877flow1,l877flow2,l877flow3}.  This
yields the rapidity, transverse momentum, and centrality dependence of
the Fourier coefficients $v_n$ (amplitude of $n$-th harmonic) in the
decomposition:
\be
E \frac{d^3 N}{d^3 p} = 
\frac{d^3N}{p_t d p_t dy d\phi}=
\frac{1}{2\pi} \frac {d^2N}{p_t d p_t dy}(1+2 v_1 \cos (\phi)
+2 v_2 \cos(2 \phi)
+...),
\label{ed3}
\ee
where the azimuthal angle $\phi$ is measured with respect to the
(true) reaction plane.

Similarly to the analysis presented in~\cite{l877flow2,l877flow3}, the
reaction plane angle is determined in four non-overlapping
pseudorapidity windows. The 'reaction plane resolution', i.e. the
accuracy with which the reaction plane orientation is determined, is
evaluated by studying the correlation between flow angles determined in
different windows.  Finally, the flow signals are corrected for the
finite reaction plane resolution.  Details of this procedure are
described in~\cite{l877flow2,l877flow3}.

\section{Results. Directed flow of light nuclei}
\label{sresu}

We present our results in Figures~\ref{fv1d} and~\ref{fv1he} for four
centrality regions, selected in accordance with transverse energy $E_T$
measured in PCal and corresponding to the values of
$\sigma_{top}(E_T)/\sigma_{geo} \approx$ 23--13\%, 13--9\%, 9--4\%, and
$<$4\% (see Fig.~4 in~\cite{l877flow2}). The value of
$\sigma_{top}(E_T)$ is obtained by an integration of $d\sigma /dE_T$
from a given value of $E_T$ to the maximal one observed, and the
geometric cross section is defined as $\sigma_{geo}=\pi
r_0^2(A^{1/3}+A^{1/3})^2=6.13$~b, for $A=197$ and $r_0$=1.2~fm.

The amplitude of the first harmonic in the Fourier decomposition of the
azimuthal distribution, $v_1(p_t)$, for deuterons is presented in
Fig.~\ref{fv1d}.  For comparison the results for protons
from~\cite{l877flow3} are shown by open symbols.  The deuteron directed
flow signal ($v_1(p_t)$) is systematically larger than that of protons,
especially at rapidities close to the projectile rapidity.  The relative
difference in deuteron and proton flow increases with centrality.  The
centrality dependence of deuteron directed flow depends on rapidity.
While at rapidities $y<2.4$ flow almost disappears at high centralities,
in the beam rapidity region it decreases very little.

At rapidities close to beam rapidity, $3.0<y<3.2$, our acceptance
permits also the determination of the mean $p_x$ of deuterons, $\mpx_d$.
To obtain this results we use the deuteron transverse momentum spectra
which were analyzed in~\cite{john97}.  At beam rapidities deuteron
spectra were found to be approximately thermal with an effective
(Boltzmann) temperature of about $80$~MeV for centralities corresponding
to our centrality regions 2--3, and with slightly lower effective
temperatures of about 50-60~MeV for the most central collisions.  Using
the deuteron spectra~\cite{john97} and $v_1(p_t)$ from Fig.~\ref{fv1d}
(centrality regions 1--2), one obtains values of $\mpx_d = v_1 \mpt_d
\approx$ 270~MeV/c, with a systematic relative error of about 20\%
(coming from uncertainties both in the spectra and in $v_1(p_t)$).  This
value of $\mpx_d$ is approximately twice the value of $\mpx_p$ for the
same centrality and rapidity regions~\cite{l877flow3}.  This observation
is in accordance with the results of Ref.~\cite{ahle98}.  The observed
absolute values of $\mpx/A\approx 130$~MeV/c are similar to the values
$\mpx/A\approx 120$-135~MeV/c~\cite{part95} observed at a beam kinetic
energy of about 1~GeV per nucleon and significantly larger than the
values measured at still lower beam energies (cf. for instance the
result of $\mpx/A\approx 60$~MeV/c~\cite{gust88} measured at a beam
energy of 0.2~GeV per nucleon).

In Fig.~\ref{fv1he}, we compare the flow signals of different light
nuclei in the beam rapidity region for different centralities of the
collision.  The flow signal increases with mass number, flow of tritons
and $^3$He is the same within error-bars.  There is a rather large
difference between proton and deuteron flow, while for nuclei with
$A\geq 2$ the mass dependence is weaker and appears to saturate.  This
is very similar to the mass dependence of flow at much lower beam
energies of 0.2~GeV per nucleon~\cite{doss87}, where the value of $\la
p_x/p_t \ra$ was also analyzed in the beam rapidity region .  Note the
very high values of $v_1$ (of the order of 0.7-0.8) for fragments with
large transverse momenta.

The error-bars shown in Figures~\ref{fv1d} and~\ref{fv1he} represent
statistical errors only. The systematic uncertainties are dominated by three
sources: 
i) Possible particle misidentification. It leads to relative errors in
$v_1$ of approximately 5\% (see above). 
ii) The uncertainty in the determination of the reaction plane
resolution~\cite{l877flow2,l877flow3}, leading to a relative error in
$v_1$ of the order of 5--10\%, similar for all particle species. 
iii) The uncertainty in correction for finite detector occupancy (see
~\cite{l877flow3}). The accuracy of the correction itself we estimate to
be of the order of 20--30\%. The occupancy correction strongly depends
on the particle transverse momentum. The correction is maximal for the
lowest $p_t$ points, where it reaches the (absolute) values of 0.1--0.12
(with an uncertainty of the order of 0.03-0.05). This large
uncertainty in the occupancy corrections in the spectrometer region
close to the beam limits our measurements at very low $p_t$.
The occupancy correction is negligible at $p_t \geq $0.6--0.8~GeV/c.

\section{Discussion}

In a simple picture where the directed flow is solely due to a common
collective motion of the matter one would expect that for light
fragments $\mpx_A=A \mpx_p$.  At lower beam energies this equality was
found to be approximately valid at rapidities close to the projectile
rapidity~\cite{gust88,gutb89}.  At lower rapidities, the ratio $\la
p_x/A \ra /\mpx_p$ was found to be {\it increasing} with particle mass.
For $A\leq 4$ the dependence is rather significant; $d(\la p_x/A
\ra)/dy$ may increase by almost a factor of 3 from protons to
$^4$He~\cite{part95,huan96}.  In our study we analyze the transverse
momentum dependence of directed flow.  Note that an equality $\mpx_A=A
\mpx_p$ does not imply that $v_1^{A}(p_t)=A v_1^{p}(p_t)$.  In the {\em
(sideward) moving thermal source} model~\cite{l877flow3,lvrd} $v_1(p_t)$
to first order does not depend on the mass of the particle, and, for
example, $v_1^{d}(p_t)=v_1^{p}(p_t)$.  Still, in the same model
$\mpx_d=2\mpx_p$, which is due to the increase of $\la p_t \ra$ with the
mass of the particle (assuming also a linear dependence of $v_1$ on
$p_t$).\footnote{ More precisely $\mpx_d/\mpx_p \approx (T_d/T_p)
(v_1^d/v_1^p)$, where $T_d$ and $T_p$ are deuteron and proton inverse
slope constants and the ratio $(v_1^d/v_1^p)$ stands for the average
ratio of deuteron and proton $v_1(p_t)$.}  Also note that in the simple
{\em coalescence} model where in order for nucleons to coalesce one
requires that they are close to each other only in momentum space,
$v_1^{A}(p_t) \approx A v_1^{p}(p_t/A)$.  Taking into account that
$v_1^{p}(p_t)$ depends almost linearly on $p_t$ one arrives once more at
the equality $v_1^{d}(p_t) \approx v_1^{p}(p_t)$.  The experimental data
(Fig.~2) show that, in the rapidity region $y<2.6$, $v_1^{d}(p_t)$ is
indeed close to $v_1^{p}(p_t)$.  However, at larger rapidities, we
observe a significant excess of deuteron flow in comparison with that of
protons.  It could imply that volume effects (to form a deuteron both
nucleons should be close to each other not only in momentum space but
also in configuration space) and/or projectile fragmentation processes
become significant in deuteron production.

In order to check if a coalescence picture including volume effects
could account for our data we have used the coalescence model of
Ref.~\cite{mati95} combined with the RQMD (version 2.3) event
generator~\cite{lrqmd}.  This coalescence model~\cite{mati95} explicitly
requires the nucleons, in order to coalesce, to be close both in the
momentum and in the configuration space.  The closeness is defined by
the cluster wave function, the parameters of which were determined by
measured root mean square charge radii of the clusters (of the order of
1.5--2.0 fm).  The results of our calculations of $v_1(p_t)^{d,p}$
corresponding to the centrality region 1 are shown in Fig.~4.  One can
see that the model describes most of the features observed in the data
(Fig.~2).  The exception is, as was already observed in proton
flow~\cite{l877flow3}, that the data exhibit an approximately linear
dependence of $v_1$ on $p_t$ at all rapidities, while the model shows a
rather fast saturation of $v_1$ in the projectile rapidity region. More
relevant for the present discussion though, the model does describe the
fact that $v_1^d(p_t)$ is very similar to $v_1^p(p_t)$ at rapidities
$y<2.6$ and that the difference becomes large at rapidities close to
beam rapidity.  Looking into details of the nucleon freeze-out
configuration space distribution (used for the coalescence) one can
notice that, for rapidities $y > 2.6$, nucleons emitted in flow
direction come from a significantly narrower spatial distribution
(especially along the flow direction) as compared to those emitted in
the opposite direction.  A smaller width of the spatial distribution
means a higher probability for nucleons to coalesce. As a consequence,
the deuteron distributions get an extra asymmetry beyond what is due to
the asymmetry in momentum space of the nucleon distributions. This may
account for the effect observed both experimentally and in the model.
 
The difference in proton and deuteron flow as a function of rapidity
could be also due to the presence of deuterons from projectile
fragmentation. In a detailed comparison~\cite{john97} of the coalescence
model~\cite{mati95} with the E877 experimental data on deuteron
production it was shown that the model fails to reproduce the deuteron
$p_t$ spectra in the projectile rapidity region, while describing the
data fairly well at smaller rapidities. It suggests that the deuteron
production mechanism at beam rapidity may not be (dominantly)
coalescence. At present a quantitative description of the fragmentation
process at the AGS energies is not available yet. However, qualitatively
one could argue that deuterons from projectile fragmentation which are
concentrated close to beam rapidity exhibit stronger flow as they are
less distorted by thermal motion, in line with the trend observed in the
data.

\section{Conclusion}

In summary, the directed flow of light nuclei has
been measured in Au+Au collisions at AGS energies in the forward
rapidity region. 
The effect has been analyzed as a function of particle
transverse momentum for different centralities of the collision.
The largest flow has been observed in collisions of approximately
half overlapping nuclei.
In such collisions and at rapidities close to  the beam
rapidity region, light nuclei exhibit very strong directed flow, 
corresponding to $v_1\approx 0.7$-0.8 at particle transverse 
momenta of about 1~GeV/c.

The directed flow ($v_1$) increases with the mass of the light nucleus.
Coalescence model calculations show that the increase in $v_1$
could be accounted for by asymmetries in the distributions of nucleons
(forming the light nucleus) both in momentum and in configuration space.

\section*{ACKNOWLEDGMENTS}

We thank the AGS staff, W. McGahern and Dr. H. Brown for excellent
support and acknowledge the help of R. Hutter in all
technical matters. Financial support from the US DoE, the NSF, the
Canadian NSERC, and CNPq Brazil is gratefully acknowledged. One of us
(JPW) thanks the A. v. Humboldt Foundation for support.


\clearpage
\newpage
\section*{Figure captions}
\begin{enumerate}

\item \label{fe877}
  The E877 apparatus.

\item \label{fv1d}
  Transverse momentum dependence of the first 
  moment ($v_1$) of the deuteron (filled circles) 
   and proton~\cite{l877flow3} (open symbols) azimuthal distributions 
  for different particle 
  rapidities and centralities of the collision.

\item \label{fv1he}
  Transverse momentum dependence of directed flow $v_1(p_t)$ of
  protons, deuterons, tritons, $^3$He, and $^4$He
  for different centralities of the collision. All particles are from
  the rapidity region $3.0<y<3.2$.
  
\item \label{fv1d_rqmd}
  Transverse momentum dependence of $v_1(p_t)$ of protons 
  and deuterons in the coalescence model combined with the RQMD 
  event generator. 
  The centrality corresponds to the experimental region 1.

\end{enumerate}



\clearpage
\newpage

\begin{figure}
\centerline{\psfig{figure=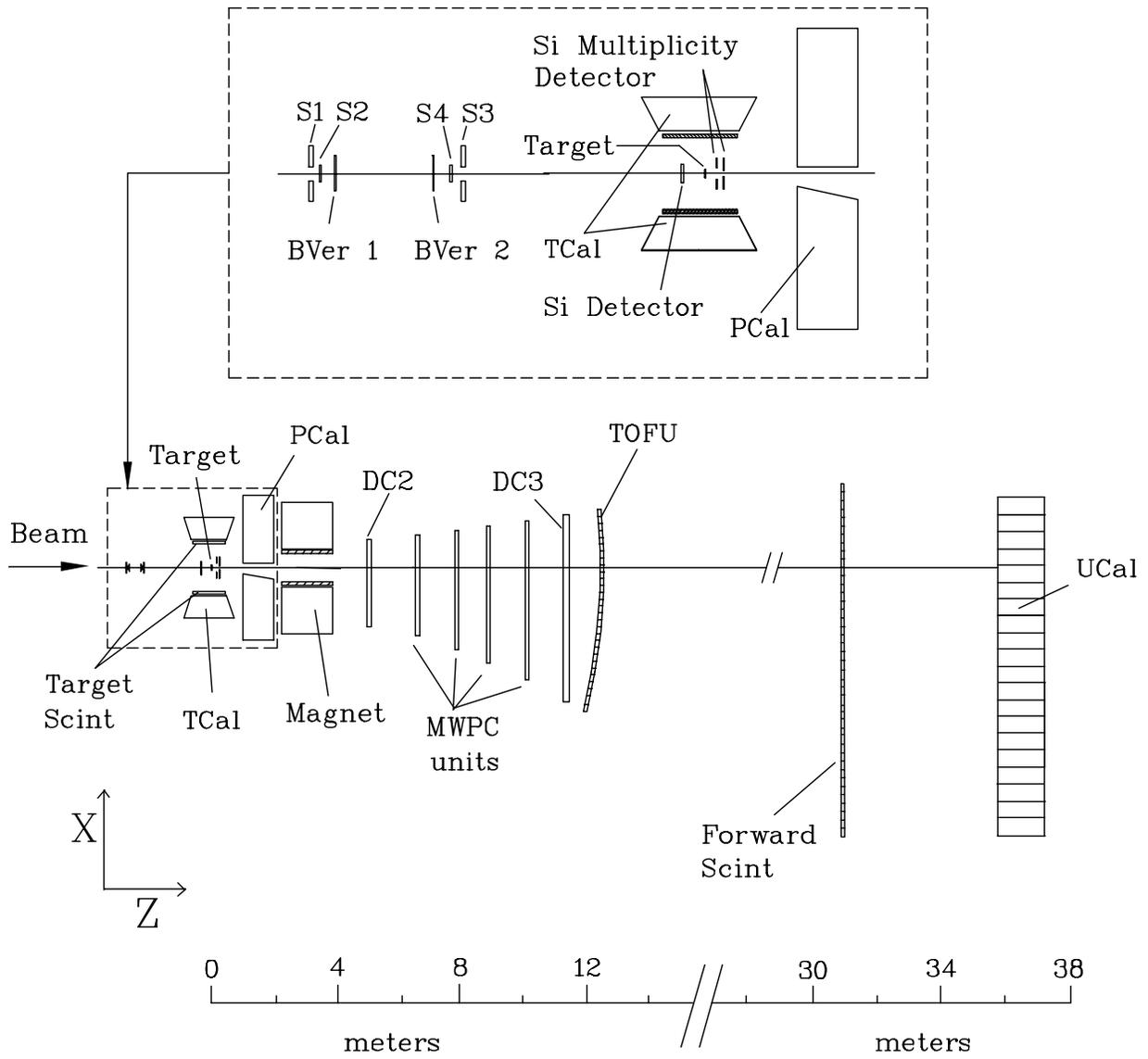,height=15.0cm}}
  \caption[]{
  The E877 apparatus.
    }
\end{figure}

\begin{figure}
\centerline{\psfig{figure=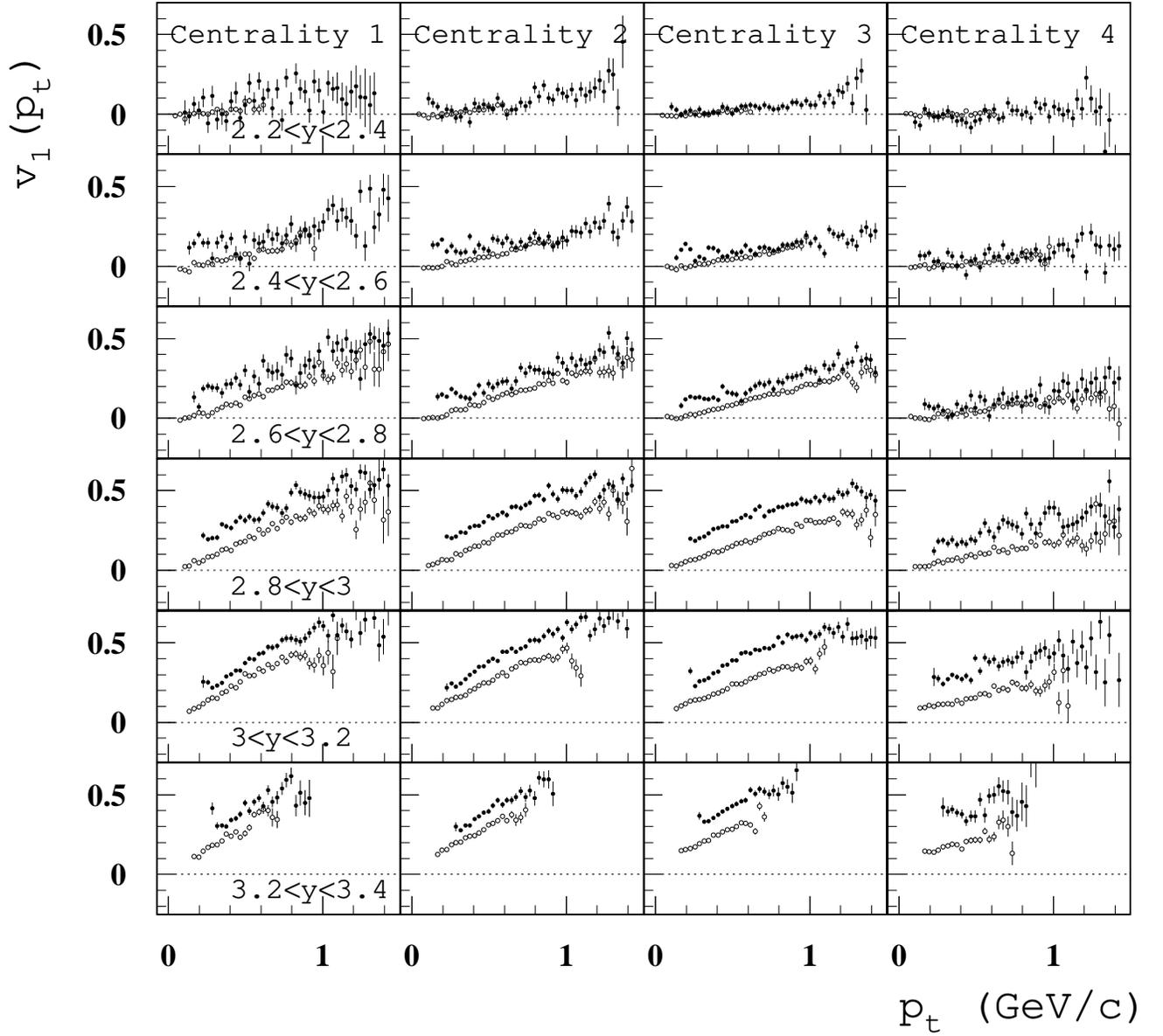,height=18.5cm}}
  \caption[]{
  Transverse momentum dependence of the first 
  moment ($v_1$) of the deuteron (filled circles) 
   and proton~\cite{l877flow3} (open symbols) azimuthal distributions 
  for different particle 
  rapidities and centralities of the collision.
    }
\end{figure}

\begin{figure}
\centerline{\psfig{figure=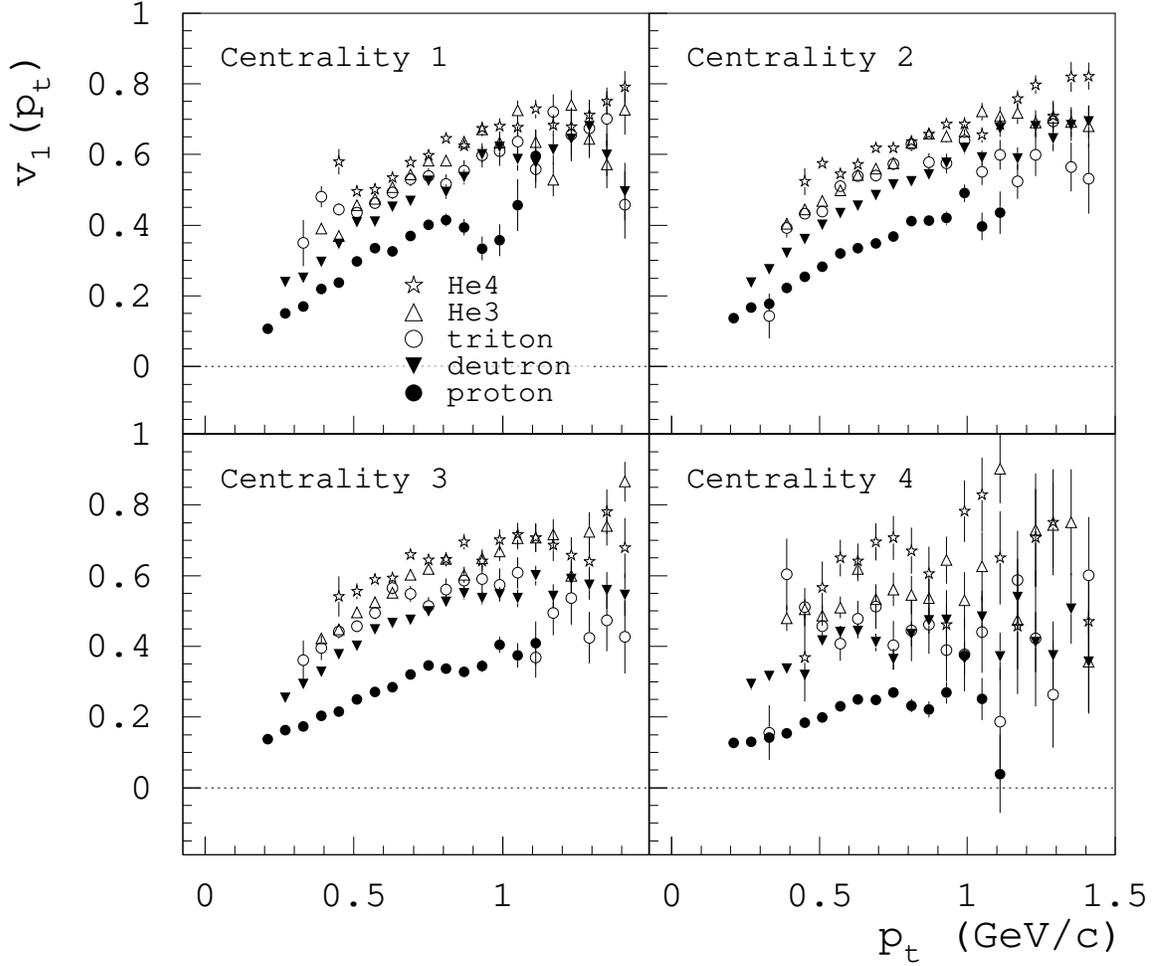,height=16.0cm}}
  \caption[]{
  Transverse momentum dependence of directed flow $v_1(p_t)$ of
  protons, deuterons, tritons, $^3$He, and $^4$He
  for different centralities of the collision. All particles are from
  the rapidity region $3.0<y<3.2$.
    }
\end{figure}
\begin{figure}

\centerline{\psfig{figure=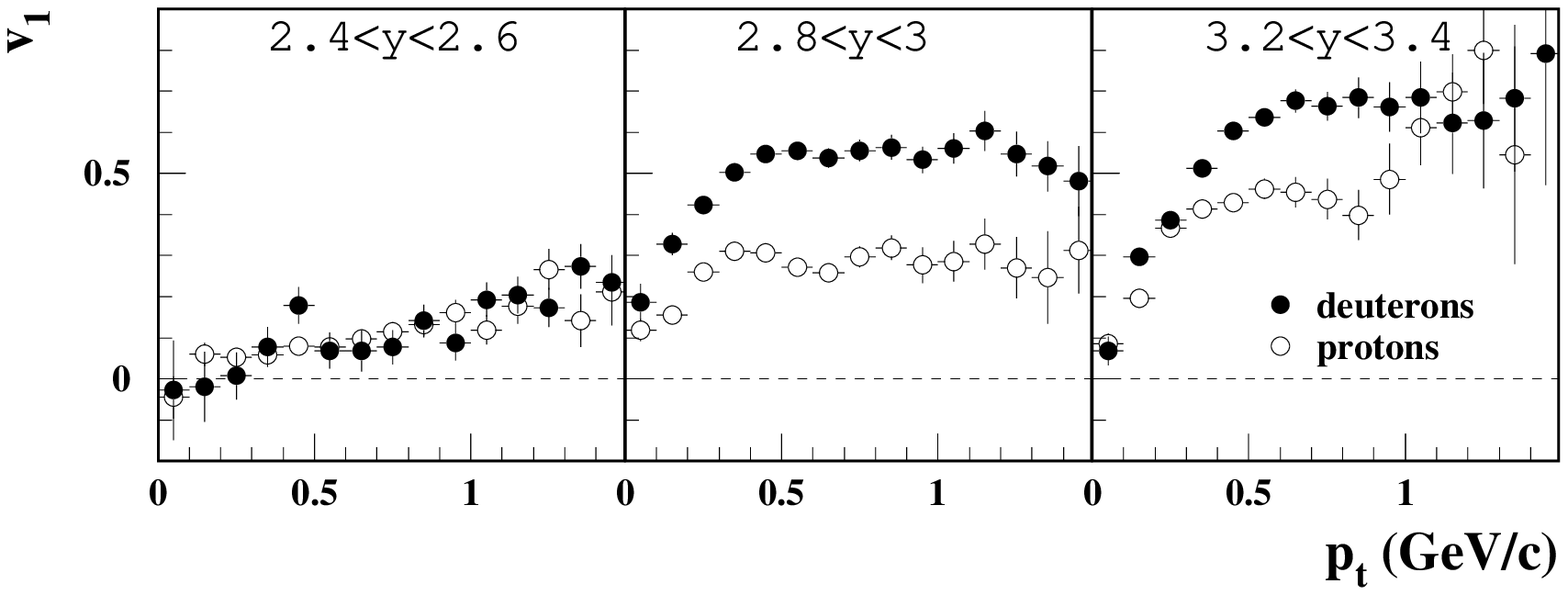,height=9.0cm}}
  \caption[]{
  Transverse momentum dependence of $v_1(p_t)$ of protons 
  and deuterons in the coalescence model combined with the RQMD 
  event generator. 
  The centrality corresponds to the experimental region 1.
    }
\end{figure}


\begin{thebibliography}{99}

\bibitem{lqm97} Proceedings of the 13th International Conference 
  on Ultra-Relativistic Nucleus-Nucleus Collisions, Quark Matter '97, 
  Nucl. Phys. {\bf A638} (1998), in print.

\bibitem{ritt98} W. Reisdorf and H.G. Ritter,
  Annu. Rev. Nucl. Part. Sci. {\bf 47}, 663 (1998).

\bibitem{gutb89} H.H. Gutbrod, A.M. Poskanzer, and H.G. Ritter,
  Rep. Prog. Phys. {\bf 52}, 1267 (1989).

\bibitem{koch90} V. Koch, B. Bl\"attel, W. Cassing, U. Mosel, and
  K. Weber, Phys. Lett. {\bf 241}, 174 (1990).

\bibitem{stoc94} G.Peilert, H. St\"ocker, and W. Greiner,
  Rep. Prog. Phys. {\bf 57}, 533 (1994).

\bibitem{mati95} R. Mattiello, A. Jahns, H. Sorge, H. St\"ocker, 
  and W.~Greiner, 
  Phys. Rev. Lett. {\bf 74} 2180 (1995);
  R.~Mattiello, H.~Sorge, H.~St\"ocker, and W.~Greiner,
  Phys. Rev. {\bf C55} 1443 (1997).

\bibitem{doss87} K.G.R. Doss {\it et al.}, 
  Phys. Rev. Lett. {\bf 59}, 2720 (1987).

\bibitem{gust88} H.\AA. Gustafsson, H.H.~Gutbrod, J.~Harris,
  B.V.~Jacak, K.H.~Kampert, B.~Kolb, A.M.~Poskanzer, H.G.Ritter,
  and H.R.~Schmidt,
  Mod. Phys. Lett. {\bf 3}, 1323 (1988).

\bibitem {ogil89} C.A. Ogilvie {\it et al.},
  Phys. Rev. {\bf C40}, 2592 (1989).

\bibitem{sull90} J.P. Sullivan {\it et al.},
  Phys. Lett. {\bf B249}, 8 (1990).

\bibitem{wang95} S. Wang {\it et al.}, 
  Phys. Rev. Lett. {\bf 74}, 2646 (1995).

\bibitem{part95} M.D. Partlan {\it et al.}, 
  Phys. Rev. Lett. {\bf 75}, 2100 (1995).

\bibitem{huan96} M.J. Huang {\it et al.}, 
  Phys. Rev. Lett. {\bf 77}, 3739 (1996)

\bibitem{cro97} P. Crochet {\it et al.}, FOPI Collaboration,
Nucl. Phys. {\bf A624}, 755 (1997).

\bibitem{dani95} P. Danielewicz, Phys. Rev. {\bf C51}, 716 (1995).

\bibitem{ahle98}  L. Ahle {\it et al.}, E802 Collaboration,
  Phys. Rev. {\bf C57}, 1416 (1998).

\bibitem{l877flow1}  J. Barrette {\it et al.}, E877 Collaboration,
  Phys. Rev. Lett. {\bf 73}, 2532 (1994).

\bibitem{l877flow2}  J. Barrette {\it et al.}, E877 Collaboration, 
  Phys. Rev. {\bf C55}, 1420 (1997).

\bibitem{l877flow3}  J. Barrette {\it et al.}, E877 Collaboration, 
  Phys. Rev. {\bf C56}, 3254 (1997).

\bibitem{lvzh} S.~Voloshin and Y.~Zhang, Z. Phys. {\bf C70}, 665 (1996);

\bibitem{lolli} J.-Y. Ollitrault, preprint nucl-ex/9711003 (1997).

\bibitem{john97} S.C. Johnson, PhD thesis, SUNY Stony Brook, 1997.  

\bibitem{lvrd} S.A. Voloshin, Phys. Rev. {\bf C55}, 1630 (1997).

\bibitem{lrqmd} H.~Sorge, A.v.~Keitz, R.~Mattiello, H.~St\"{o}cker,
       and W.~Greiner, Phys. Lett. {\bf B243}, 7 (1990);
  H. Sorge, in Proceedings of the Workshop "Heavy-Ion Physics at the AGS '96," 
  eds. C.A. Pruneau, G. Welke, R. Bellwied, S.J. Bennett, 
  J.F. Hall, and W.K. Wilson, Wayne State University, 1996.

\end{thebibliography}
\end{document}